\documentclass[aps,pre,preprint,groupedaddress,amsmath,amssymb,showpacs,preprintnumbers]{revtex4}
\providecommand{\mean}[1]{\langle #1 \rangle}
\usepackage{graphicx,color}

\begin{document}

\title{Exact moments in a continuous time random walk \\ with complete memory of its history}
\author{Francis N.~C.~Paraan}
\email{fcparaan@up.edu.ph}
\author{J.~P.~Esguerra}
\email{pesguerra@nip.upd.edu.ph}
\affiliation{National Institute of Physics, University of the Philippines, Diliman, Quezon City}
\date[]{Received 21 March 2006}
\preprint{PREPRINT: TPG270506}

\begin{abstract}
We present a continuous time generalization of a random walk with complete memory of its history [Phys. Rev. E {\bfseries 70}, 045101(R) (2004)] and derive exact expressions for the first four moments of the distribution of displacement when the number of steps is Poisson distributed. We analyze the asymptotic behavior of the normalized third and fourth cumulants and identify new transitions in a parameter regime where the random walk exhibits superdiffusion.  These transitions, which are also present in the discrete time case, arise from the memory of the process and are not reproduced by Fokker-Planck approximations to the evolution equation of this random walk.
\end{abstract}


\maketitle

\section{Introduction}\label{intro}
Recently, Sch\"{u}tz and Trimper obtained the exact moments of the probability distribution function (PDF) of total displacement of a discrete time random walk with memory of its entire history \cite{schutz}. They coined the term ``elephant random walk'' (ERW) for this model, referring to the myth that elephants have perfect memories. In the ERW, long-range memory effects enter the system dynamics through the microscopic evolution equation as a displacement- and time-dependent bias.  This microscopic approach to modeling memory contrasts sharply with the largely established practices of introducing correlated random noise terms in Langevin equations \cite{hanggi}, memory kernels in generalized Fokker-Planck equations \cite{metzler, trimper}, or combinations of both \cite{lutz,bazzani} at a coarse-grained scale.

In its simplest formulation, the discrete time ERW takes place on a one-dimensional lattice with lattice constant $\Delta$ and each step occurs at time intervals $1/\lambda$. The memory of this process is controlled by the initial bias $\beta = 2q-1$ and the correlation parameter $\alpha = 2p-1$. Here, $q$ is the probability of taking the first step in the positive direction while $p$ is the probability of taking succeeding steps in the same direction as a randomly chosen displacement in the walker's history. The corresponding microscopic evolution equation for the probability $P_{N+1}(y)$ of the walker having a total displacement $y$ from the origin after $N+1$ steps is therefore
\begin{equation}\label{evol}
\begin{split}
P_1(y) &= \frac{1}{2}(1-\beta)\delta(y+\Delta) + \frac{1}{2}(1+\beta)\delta(y-\Delta), \\
P_{N+1}(y) &= \frac{1}{2}\left[ 1-\frac{\alpha(y+\Delta)}{N\Delta}\right]	P_{N}(y + \Delta) \\ 
					 &\ + \frac{1}{2}\left[ 1+\frac{\alpha(y-\Delta)}{N\Delta}\right]	P_{N}(y - \Delta),\ N\ge1, 		 
\end{split}	
\end{equation}
where $\delta(y)$ is the usual Dirac delta distribution. Details of the stochastic rules governing the movement of the elephant random walker may be found in \cite{schutz}.

The sequence of displacements in an ERW can also be pictured as a sequence of binary events.  Such a perspective would be of great use in the study of two-level systems where successive occupation probabilities are linear functions of the fraction of occupants in the respective levels. Some examples are two queues with arrivals biased toward the shorter one and two competing states or agencies with recruitment favoring the more popular one.  Such an interpretation has been successful in the study of correlated bit strings observed in reduced sequences of DNA bases in genomes, letters in written text, and losses and gains in the stock market \cite{hod,usatenko}.

An analysis of the scaling properties of the mean and mean squared displacement of the PDF of total displacement reveals two distinct transitions in the dynamics of the walker \cite{schutz}: 
\begin{enumerate}
	\item When the parameter $\alpha$ is negative, the sequence of displacements is anticorrelated so that the walker is localized about the origin. Conversely, positive correlations lead to a net drift in the direction of the initial bias when $\alpha>0$. An uncorrelated random walk is recovered when $\alpha$ is exactly zero.
	\item When $\alpha$ is less than $1/2$, the mean squared displacement scales linearly with the number of steps $N$ as in a normal diffusive process. On the other hand, when $\alpha>1/2$, the mean squared displacement scales asymptotically as a power law that depends on $\alpha$ and results in superdiffusion. When $\alpha$ is exactly 1/2, the process is marginally superdiffusive and the second moment behaves as $\sim\Delta^2 N\ln N$ for large $N$.
\end{enumerate}

When the discrete microscopic evolution equation (\ref{evol}) is approximated by a Fokker-Planck equation in the continuum limit, the elephant random walk can be shown to reproduce the dynamics of a Brownian particle in a time-varying harmonic oscillator potential \cite{schutz}.  This system is encountered in the study of anomalous diffusion that results from time-dependent Fokker-Planck drift coefficients \cite{malacarne, lillo}. In this continuum approximation, an initially sharply peaked PDF of displacement asymptotically evolves into a Gaussian distribution with a time-dependent envelope.  

This Brief Report has two main contributions.  First, we present a continuous time generalization of the ERW by taking the time intervals between steps from a distribution for which the central limit theorem applies (Sect.\ \ref{model}). This generalization leads to a more flexible model as the waiting time between successive displacements (or events) in the ERW is no longer required to be constant.  We will focus on an exponential waiting time distribution in a Poisson generalization that will allow us to obtain exact expressions for the moments of the distribution of displacement in terms of confluent hypergeometric functions (Sect.\ \ref{poisson}). Second, we calculate the normalized third and fourth cumulants, or the coefficients of skewness and kurtosis, of the PDF of displacement in an attempt to discover new dynamical transitions and to analyze the range of validity of the continuum Fokker-Planck approximation described above (Sect.\ \ref{coeff}).

\section{The Continuous Time ERW}\label{model}
We consider an elephant random walk in continuous time with a waiting time distribution $\psi(t)$ that satisfies the following conditions: $(i)$ This distribution represents mutually independent random waiting times, and $(ii)$ It possesses finite first and second moments. For such a distribution, the probability density $\mathcal{P}(N,t)$ that $N$ steps will take place in the ERW at time $t$ is given by \cite{montroll}
\begin{equation}
	\mathcal{P}(N,t) = \bigl[ \psi(t) \, \ast \, \bigr]^N \left[ 1-\int_0^t \psi(t')\, dt' \right],
\end{equation}
where the asterisk $\ast$ is the Laplace convolution operator. Formally, the continuous time PDF of total displacement $P(y,t)$ is therefore
\begin{equation}
	P(y,t) = \sum_{N=0}^{\infty}\mathcal{P}(N,t)P_N(y),
\end{equation}
and its moments $\mean{y_t^m}$ are 
\begin{equation}
	\mean{y_t^m} =  \sum_{N=0}^\infty \mathcal{P}(N,t) \mean {y_N^m},
\end{equation}
where $\mean{y_N^m}$ are the moments of the $N$ step PDF of displacement $P_N(y)$ in discrete time. The central limit theorem \cite{feller} guarantees that in the asymptotic limit the continuous time PDF $P(y,t)$ becomes identical to the analogous distribution of the discrete time ERW having the same average waiting time. That is, if $1/\lambda$ is the mean waiting time, then asymptotically
\begin{equation}\label{correspondence}
	P(y,t \approx N/\lambda) \approx P_N(y) \quad \text{as} \quad N \rightarrow \infty.
\end{equation}
Thus, all of the dynamical transitions observed in the discrete time ERW are also present in our continuous time generalization and occur at the same critical values. Furthermore, the same continuum approximations used in \cite{schutz} may also be applied to this continuous time model with the same degree of success.

From this well-known result, we are able to obtain exact closed form results for the moments of $P(y,t)$ if we consider a waiting time distribution given by the exponential 
\begin{equation}
	\psi(t) = \lambda \, e^{-\lambda t}.
\end{equation}
The number density $\mathcal{P}(N,t)$ is now given by the Poisson distribution and the moments of $P(y,t)$ are 
\begin{equation}\label{momformula}
	\mean{y_t^m} =  e^{-\lambda t} \sum_{N=0}^\infty \frac{(\lambda t)^N}{N!} \mean {y_N^m}.
\end{equation}
In the following section, we shall calculate the first four moments of $P(y,t)$ in this Poisson ERW and show that they indeed approach the analogous discrete time moments as demanded by the central limit theorem (\ref{correspondence}).

\section{Moments in the Poisson ERW}\label{poisson}
To calculate the moments of $P(y,t)$, we will need the exact expressions for the moments of the displacement in a discrete time ERW and substitute these values into (\ref{momformula}). The first two discrete time moments are given in \cite{schutz} as well as a general recursion formula that will allow us to calculate the higher moments. Quoting these results and solving for the third and fourth moments in an $N$ step discrete time ERW yields
\begin{align}
\mean{y_N^1} &= \frac{\beta\Delta}{\alpha}N\frac{(\alpha)_N}{(1)_N}, \\ 
\mean{y_N^2} &= \frac{\Delta^2}{2\alpha-1}\left[N \frac{(2\alpha)_N}{(1)_N} - N \right], \\
\mean{y_N^3} &= \frac{\beta\Delta^3}{\alpha(2\alpha-1)}\biggl\{ \Bigl[\left(\alpha+1\right)N\Bigr]\frac{(3\alpha)_N}{(1)_N}\nonumber\\ 
						 & \quad - \Bigl[ 3N^2+(\alpha+1)N \Bigr]\frac{(\alpha)_N}{(1)_N}\biggr\},	\\
\mean{y_N^4} &= \frac{\Delta^4}{(2\alpha-1)^2}\biggl\{ \left[ \frac{6(2\alpha^2+2\alpha-1)}{4\alpha-1} N \right] \frac{(4\alpha)_N}{(1)_N} \nonumber \\
             & \quad - \Bigl[6N^2+4(\alpha+1)N\Bigr]\frac{(2\alpha)_N}{(1)_N} \nonumber \\
             & \quad + 3N^2+\frac{2(2\alpha^2+1)}{4\alpha-1}N  \biggr\}.						 
\end{align}
Here, we have made extensive use of the rising factorial or Pochhammer symbols $(a)_N$ \footnote{The rising factorial or Pochhammer symbols are defined as $(a)_N = a(a+1)(a+2)\dotsm(a+N-1)$, with $(a)_0 = 1$.} in anticipation of encountering hypergeometric sums as we solve for the moments of the PDF of displacement. 

To illustrate the evaluation of the continuous time moments, we shall calculate the mean total displacement explicitly. By making the appropriate substitution into (\ref{momformula}) and considering the action of the $ \lambda t\, \partial/\partial (\lambda t)$ operator, we are led to the expression
\begin{equation}
\mean{y_t^1} = \frac{\beta\Delta e^{-\lambda t}}{\alpha} \lambda t \frac{\partial}{\partial(\lambda t)} \sum_{N=0}^\infty  \frac{(\lambda t)^N}{N!} \frac{(\alpha)_N}{(1)_N}.
\end{equation}
Recognizing the power series representation of Kummer's confluent hypergeometric function $M(a,c;\lambda t)$ \footnote{$M(a,c;z) = \sum_{n=0}^\infty \frac{z^n}{n!}\frac{(a)_n}{(c)_n}$.} and performing the differentiation yield the following exact result:
\begin{equation}\label{one}
\mean{y_t^1} = \beta\Delta \lambda t \, e^{-\lambda t}M(\alpha + 1, 2; \lambda t).
\end{equation}

Similarly, we obtain the following expressions for the second, third, and fourth moments of the distribution of total displacement:
\begin{align}\label{two}
\mean{y_t^2}&= \frac{\Delta^2 \lambda t \, e^{-\lambda t} }{2\alpha-1}\Bigl[2\alpha M(2\alpha+1,2;\lambda t) - e^{\lambda t}  \Bigr], \\
\mean{y_t^3} &= \frac{\beta\Delta^3 \lambda t \, e^{-\lambda t}}{2(2\alpha-1)} \Bigl[ 6( \alpha+1 ) M(3\alpha+1,2;\lambda t) 
\nonumber \\ & \quad - 2(\alpha+4)M(\alpha+1,2;\lambda t) \Bigr.
\nonumber \\ & \quad - 3(\alpha+1)\lambda t\, M(\alpha+2,3;\lambda t)  \Bigr], \label{three} \\
\mean{y_t^4} &= \frac{\Delta^4 \lambda t \, e^{-\lambda t}}{(2\alpha-1)^2} \biggl\{ \frac{24\alpha( 2\alpha^2 + 2\alpha -1 )}{4\alpha-1} M(4\alpha+1,2;\lambda t) 
\nonumber \\ & \quad -4\alpha(2\alpha+5)M(2\alpha+1,2;\lambda t) \biggr. \nonumber \\ 
						 & \quad - 6\alpha \bigl(2\alpha+1\bigr) \lambda t\, M(2\alpha+2,3;\lambda t) \biggr. \nonumber \\ 
						 & \quad + e^{\lambda t}\left[3\lambda t + \frac{4\alpha^2+12\alpha-1}{4\alpha-1} \right] \biggr\}.
\label{four}
\end{align}

Applying L'H\^opital's rule as the parameter $\alpha$ is made to approach $1/2$ allows us to find the scaling properties of these moments when the process is marginally superdiffusive. As in the discrete time process, we find that the second, third, and fourth moments behave asymptotically as $\Delta^2 \lambda t \ln \lambda t$, $6\pi^{-1/2}\beta\Delta^3 (\lambda t)^{3/2} \ln \lambda t$, and $3\Delta^4 (\lambda t \ln \lambda t)^2$, respectively. At the critical point $\alpha = 1/4$, a mild transition arising from the fourth moment is observed for small times but is obscured in the asymptotic limit.

\begin{figure}[ht]
  \begin{center}
   \includegraphics[bb = 0 695 592 841, width=1\linewidth,angle=0]{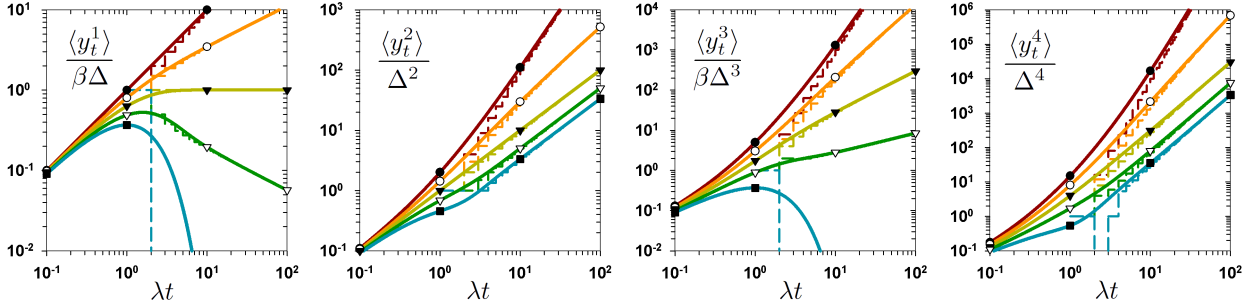}
  \end{center}
\caption{(Color online) The moments of the total displacement in a Poisson continuous time elephant random walk (solid lines) asymptotically coincide with the moments of a discrete time ERW (dashed lines) with the same mean waiting time $1/\lambda$. Representative curves have the memory parameter $\alpha$ equal to $1.0\ (\bullet)$, $0.499\ (\circ)$, $0.0\ (\blacktriangledown)$, $-0.5\ (\triangledown)$, and $-1.0\ (\blacksquare)$.  Note that the odd moments vanish when the initial bias $\beta$ is zero.\label{12mom}}
\end{figure}

From the asymptotic series expansion of the confluent hypergeometric function \cite{stegun} we find that the leading terms in the asymptotic expansions of these continuous time moments are identical to that of the corresponding discrete time moments for $\lambda t \approx N$, as we asserted earlier. In Fig.\ \ref{12mom} we confirm graphically that the dimensionless moments in a Poisson continuous time ERW approaches those in a discrete time ERW with the same mean waiting time. In these plots we have used several values of $\alpha$ to illustrate the behavior of these moments in several regimes of the ERW: superdiffusive ($\alpha=1.0$), correlated diffusive ($\alpha=0.499$), no memory ($\alpha=0.0$), and anticorrelated diffusive ($\alpha=-0.5, -1.0$). We find good agreement between corresponding moments as early as when $t \approx 100/\lambda$.

\section{Coefficients of Skewness and Kurtosis}\label{coeff}
In this section, we calculate the coefficients of skewness and kurtosis and examine their scaling properties in the long time limit. Solving for these quantities will yield relative measures of the skewness and kurtosis of the PDF in relation to its width and will allow us to identify dynamical transitions in these quantities. If we label the time dependent cumulants of the PDF of displacement as $\kappa_t^m$, the coefficients of skewness and kurtosis are defined as $c_t^3 = {\kappa_t^3}/{(\kappa_t^2)^{3/2}}$ and $c_t^4 = {\kappa_t^4}/{(\kappa_t^2)^2}$, respectively. 

\begin{figure}[t]
  \begin{center}
    \includegraphics[bb = 0 587 593 841, width=1\linewidth,angle=0,clip]{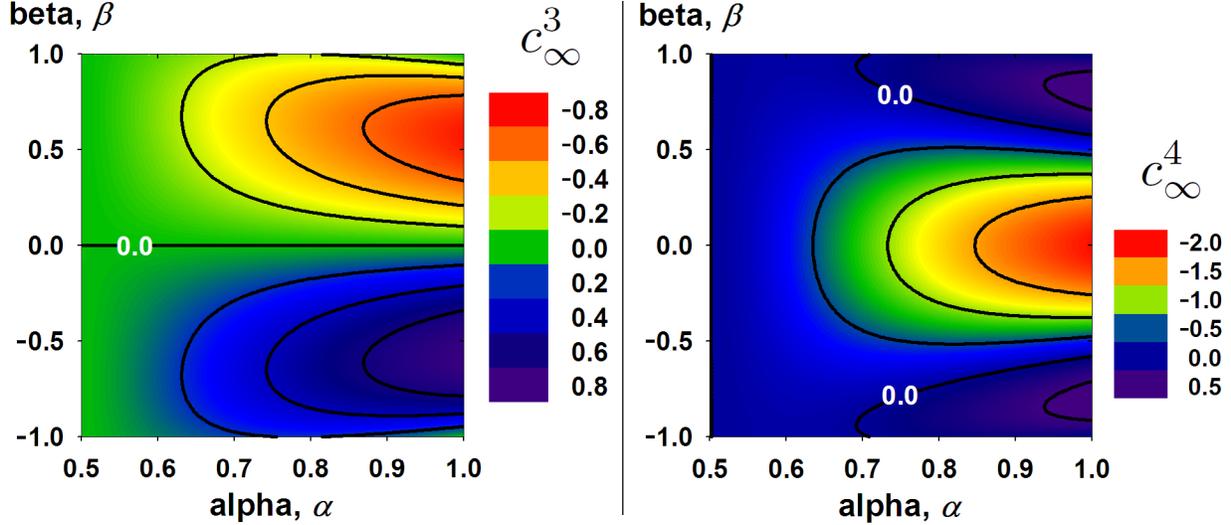}
  \end{center}
\caption{(Color online) Asymptotic values of the coefficient of skewness ({left}) and kurtosis ({right}) in the superdiffusive regime of an elephant random walk. \label{skewkurt}}
\end{figure}

By considering the leading order terms of the asymptotic expansions of these cumulants and taking the limiting value of $c_t^3$ and $c_t^4$ as $t/\lambda$ approaches infinity, we find that these coefficients vanish asymptotically when $\alpha < 1/2$. This long time behavior under the condition of normal diffusive scaling is correctly reproduced by the continuum Gaussian approximations discussed earlier and justifies its use in the diffusive regime. However, when $\alpha > 1/2$ the coefficients asymptotically approach constant values and the PDF of displacement is no longer approximately Gaussian. In this regime of superdiffusion the normalized coefficients depend asymptotically on the memory parameters $\alpha$ and $\beta$ according to
\begin{align}
	c_\infty^3 & = -\beta\sqrt{2\alpha-1}\bigl[\Gamma(2\alpha)\bigr]^{3/2} \biggl\{
 \frac{3}{\Gamma(2\alpha)\Gamma(\alpha+1)} \nonumber \\
						 & \quad - \frac{3(\alpha+1)}{\Gamma(3\alpha+1)} - \frac{2\beta^2(2\alpha-1)}{[\Gamma(\alpha+1)]^3} \biggr\}, \\
  c_\infty^4 & = -3 + 6\bigl[ \Gamma(2\alpha)\bigr]^2 \biggl\{ \frac{(2\alpha^2+2\alpha-1)}{(4\alpha-1)\Gamma(4\alpha)} \nonumber \\
  					 & \quad - \frac{2\beta^2(2\alpha^2 +\alpha-1)}{\Gamma(3\alpha+1)\Gamma(\alpha+1)} + \frac{2\beta^2(2\alpha-1)}{\Gamma(2\alpha)[\Gamma(\alpha+1)]^2} \nonumber \\
  					 & \quad - \frac{\beta^4(2\alpha-1)^2}{[\Gamma(\alpha+1)]^4}\biggr\}. 						 
\end{align}
These asymptotic values are plotted in Fig.\ \ref{skewkurt} where we have labeled the critical contours where these coefficients vanish.

We find that the skewness of the PDF has the opposite sign as the initial bias $\beta$. Due to strong positive correlations in the superdiffusive regime, taking steps in the direction of the initial bias becomes even more favorable as the walk progresses. Thus, when the walk is initially biased in the positive direction, the most probable displacement becomes greater than the mean displacement at large times and the PDF becomes negatively skewed. 

There are two transition curves separating asymptotically platykurtic ($c_\infty^4 < 0$) and leptokurtic ($c_\infty^4 > 0$) PDFs shown in Fig.\ \ref{skewkurt}. The peak of the PDF of displacement becomes relatively sharper than a similar Gaussian distribution as the dynamics of the walker becomes ballistic when $\alpha$ approaches unity and $\beta$ approaches $\pm 1$. Indeed, recent numerical calculations of the PDF of displacement in superdiffusive ERWs with bias $\beta =1$ \cite{dasilva} indicate the presence of such deviations from a Gaussian distribution in the critical central region. The opposite behavior is observed as the initial bias $\beta$ vanishes. We find that in the superdiffusive regime the memory effects result in a flatter distribution of displacement than the corresponding distribution of the memory-less Bernoulli walk \cite{hughes}.

The general non-vanishing skewness and kurtosis of the PDF and the additional memory-induced transitions in the superdiffusive regime that we have described above are not observed in the time-dependent Gaussian approximation obtained from a Fokker-Planck equation. The origin of the breakdown of this approximation when $\alpha > 1/2$ becomes clear if we consider a theorem by Pawula \cite{pawula} that states that a necessary condition for the justified truncation of a Kramers-Moyal expansion of the microscopic evolution equation is that normalized cumulants beyond the second should vanish \cite{risken}. This condition is clearly not satisfied in superdiffusive elephant random walks.  Although a Fokker-Planck approximation to the evolution equation will yield correct asymptotic mean and mean squared displacements, nontrivial information on the PDF of displacement that is related to higher non-vanishing cumulants is lost.

\section{Conclusion}

In this paper, we present a continuous time random walk model with complete memory of its history and calculate the exact moments of the distribution of displacement for the case of an exponentially decaying waiting time distribution. In this Poisson generalization, we show that the continuous time moments are asymptotically equal to the moments of the analogous discrete time process with the same mean waiting time as required by the central limit theorem. An analysis of the normalized third and fourth cumulants of the PDF of total displacement reveals new memory-induced transitions that appear in a parameter regime where this process exhibits anomalous diffusion ($\alpha>1/2$). The facts that these cumulants do not vanish and that these transitions exist show that Fokker-Planck approximations are inadequate in the superdiffusive regime of an elephant random walk.

\bibliography{elebib}

\end{document}